\documentclass[dvips]{acta}
\usepackage{supertabular,lscape,epsfig}
\usepackage{amssymb}
\usepackage{amsmath}
\usepackage[T1]{fontenc}

\SetPages{0}{0}

\SetVol{76}{2026}

\usepackage{lmodern}

\newcommand{\NS}{N_{\rm s}}
\newcommand{\NC}{N_{\rm in,r}}
\newcommand{\NT}{N_{\rm mech}}
\newcommand{\WS}{W_{\rm s}}
\newcommand{\WC}{W_{\rm c}}
\newcommand{\WT}{W_{\rm t}}

\newcommand{\PHS}{B_{\rm surf}}
\newcommand{\PHB}{B_{\rm bot}}
\newcommand{\RES}{\Re(\sigma)}

\newcommand{\PTOT}{\mathcal{P}}
\newcommand{\EKIN}{\mathcal{E}_{\rm kin}}
\newcommand{\PEX}{\mathcal{P}_{\rm ex}}
\newcommand{\PRSP}{\mathcal{P}_{\rm rsp}}
\newcommand{\PDB}{\Delta B}

\newcommand{\mFig}[1]{Fig.~\ref{fig:#1}}
\newcommand{\mEq}[1]{Eq.~(\ref{eq:#1})}

\DeclareMathOperator{\sgn}{sgn}

\begin{document}

\begin{Titlepage}

\Title{A generalized quadratic balance relation for nonradial nonadiabatic pulsations}

\Author{Zalewski, J.,}
{Independent researcher \\
e-mail: jan.zalewski.a2@gmail.com}

\end{Titlepage}

\Abstract{
We derive a generalized quadratic balance relation for linear nonadiabatic, nonradial stellar pulsations, starting from a sesquilinear amplitude analogue of the pressure-volume-rate work. The derivation requires neither weak nonadiabaticity nor averaging over a pulsation cycle. The relation is expressed in terms of work, generalized norm and boundary contributions and is subsequently recast into kinetic-energy-power form
\[
\PTOT+\PDB=2\RES\EKIN.
\]
The volume power is decomposed into a thermodynamic exchange term and response terms associated with compression, horizontal-area deformation and gravitational stratification effects. The equivalent forms of the relation provide diagnostics for checking the computed eigenfrequencies and estimating mode excitation rates.

The properties of the balance relation for various types of modes in an envelope of a  model AGB star are examined. We analyze the terms entering the power $\PTOT$ for radial and nonradial p-modes, strange modes as well as examples of low frequency outer-envelope gravity modes and thermal modes. The results show that the magnitude of mode driving is not determined solely by the thermodynamic work term.
}
{stars: AGB and post-AGB, stars: oscillations, stars: interiors, methods: analytical, methods: numerical}

\section{Introduction}
Since the early development of stellar pulsation theory by Eddington (1926) the work integral has provided the principal link between the local thermodynamic processes operating in a star and the excitation or damping of pulsations. In the classical interpretation the work over a pulsation cycle gives the change of pulsation energy, while the work integrand identifies the layers in which driving or damping occurs. Relating this work to the total kinetic energy of the mode provides an estimate of the excitation or damping rate (Cox 1980, Unno et al. 1989).
Buchler and Regev (1982) extended this interpretation using a multiple-time formalism applicable to slowly evolving pulsations. The classical formulation is physically transparent and highly effective for weakly nonadiabatic modes whose amplitudes change little over one oscillation period. Gautschy and Glatzel (1990), however, pointed out that for highly nonadiabatic pulsations the amplitude may change appreciably during a nominal cycle, causing the quasiadiabatic, cycle-averaged work integral to lose its usual interpretation. This difficulty becomes particularly important when the growth or damping time is comparable to the pulsation period, while a cycle-based construction is altogether unavailable for non-oscillatory thermal modes. These limitations motivate the formulation of a balance relation directly in terms of complex pulsation amplitudes, without requiring the existence of an approximately periodic pulsation cycle.

To remove the dependence on pulsation-cycle average, Glatzel (1994) developed a work-balance formulation for complex stellar pulsation eigenfunctions using an ensemble average over the arbitrary initial phase. Zalewski (2026c) adopted a closely related sesquilinear amplitude form and applied it to radial nonadiabatic pulsations, decomposing the resulting relation into work, norm, and boundary terms. In the present article, this formulation is extended to nonradial pulsations without invoking Cowling approximation in the derivation. The nonradial geometry introduces horizontal-inertia and horizontal area deformation terms, while perturbations of the gravitational potential give rise to an additional gravitational-stratification contribution and associated boundary terms. 

The resulting balance relation is written in terms of total power of the mode, the boundary terms and mode kinetic energy. The obtained formula is then used in the analysis of terms contributing to mode stability for envelope p-modes, strange modes, low frequency outer-envelope gravity modes trapped in the convection region and thermal modes for radial and nonradial oscillations.

\section{Pulsation equations}
The linearized pulsation equations which will be used in derivation of the balance equation are presented in Appendix A. The derivation follows Dziembowski (1977). We will use linearized continuity, momentum and Poisson equations in order to obtain formulation suitable for both radial and nonradial nonadiabatic pulsation.

In what follows the dependent variables used are the radial and horizontal displacement perturbations, $d$ and $h$ respectively, the Lagrangian pressure ($p$) and entropy ($s$) perturbations, together with gravitational potential perturbations resulting from Poisson equation - $w$ and $w_1$ (see Appendix A).

We will also use a derived perturbation quantity
\[
q=\frac{\Delta V}{V}=-\frac{\Delta\rho}{\rho}.
\]
The independent variable is defined as $x=\ln(r/R_{\odot})$ and we adopt eigenfunctions' normalization of $d(x_{s})=1$ at the surface.

The pulsation equations which will be used in the derivation of the balance equation will be the continuity \mEq{ContQ} and radial momentum \mEq{RM}, with the horizontal displacement ($h$) obtained from \mEq{HM}.

\section{Generalized pressure-volume-rate work}
Following the formulation of Glatzel (1994) we use the linearized continuity and momentum equations to obtain a balance relation. In Zalewski (2026c) a sesquilinear form representing an analogue of the $-PdV$ pressure work was introduced, as means of avoiding the necessity for cycle-averaging 
\[{\cal P}_{\rm s}=-p\,\overline{\dot{q}}.\]
This form will be used also here for the case of nonradial pulsation.
By using the definitions given in Appendix A the time derivative of relative volume perturbation may be expressed as $\dot{q}=\sqrt{4\pi G\langle\rho\rangle}\sigma q$ to obtain the complex work kernel
\begin{equation}
	{\cal K}_s=-p\,\overline{\sigma q},
\label{eq:Ks}
\end{equation}
where the scaling factor is included in $C(x)$ defined below.

We use this kernel to define amplitude-level work contribution as
\begin{equation}
\frac{1}{2}\Re({\cal K}_s).
\label{eq:KS}
\end{equation}
The luminosity-scaled work integral is obtained from \mEq{KS} as follows. By writing
\[
\frac{dW_s}{dx}=C(x)\Re({\cal K}_s),
\]
where the factor $1/2$ is included in $C(x)$:
\[
C(x)=\frac{1}{2}4\pi r^3P\frac{\sqrt{4\pi G\langle\rho\rangle}}{L},
\]
which, using our notation can be written as
\[
C(x)=\frac{1}{2}A_{13}\frac{A_8A_{10}}{-A_7},
\]
with
\[
A_{13}=\frac{\sqrt{4\pi G\langle\rho\rangle}}{L}\frac{4\pi r^3P(-A_7)}{\nabla_{ad}},
\]
and the remaining coefficients defined in Appendix A.

With this notation the formula for the work integral is given by
\[
W_s=\int_{x_b}^{x_s}C(x)\Re({\cal K}_s)dx.
\]

Using the introduced dependent variables the work integral $W_s$ may be written as
\begin{equation}
W_s=-\int_{x_b}^{x_s} C(x)\Re(p\,\overline{\sigma q}).
\label{eq:WSQ}
\end{equation}

The work integral depends on the pressure and volume perturbation and on the phase of the pressure term relative to the rate of volume change scaled by the nondimensional factor $C(x)$. The quantities $x_b$ and $x_s$ specify the location of the inner and outer boundary.

The volume perturbation may be expressed in terms of $p$ and $s$ and hence ${\cal K}_s$ written as
\[
{\cal K}_s=-p\,\overline{\sigma q}=A_4p\,\overline{\sigma p}+A_7p\,\overline{\sigma s},
\]
leading to the following form for $W_s$
\begin{equation}
	W_s=W_c+W_t,
\label{eq:WS}	
\end{equation}
with the two terms given by
\[
W_c=\int_{x_b}^{x_s} C(x)A_4\Re(p\,\overline{\sigma p})dx,
\]
and
\[
W_t=\int_{x_b}^{x_s} C(x)A_7\Re(p\,\overline{\sigma s})dx.
\]

The second term is the negative of the classical dissipation integral containing thermodynamic term $p\,\overline{\sigma s}$, but the full work integral contains also the first, compressional term, $W_c$. This term vanishes for purely adiabatic pulsation ($\RES=0$).

Both of these terms are found to play a substantial role for radial nonadiabatic pulsation in supergiant envelopes, where it turns out that the compressional term becomes more important for the balance than the classical one even for ordinary p-modes. Both terms play an important role also for strange modes where they both have comparable but opposite contribution to the $\WS$. This is discussed in Zalewski (2026c).

Thus when dealing with nonadiabatic modes one should consider both terms in \mEq{WS}, not only the classical $W_t$ term.

It may be noted that work integral does not depend explicitly on terms associated with nonradial pulsations ($\Lambda$ or $\eta$), nor does it depend on the potential perturbations ($w$). The presented derivation also does not depend on any assumptions related to the degree of nonadiabaticity, hence it is applicable to fully nonadiabatic environments like envelopes of AGB stars, but also to nearly adiabatic pulsation.

\section{Transformation of the work form}
We will now derive the balance relation for nonadiabatic pulsations. Contrary to classical approach (see Cox 1980, Unno et al. 1989) we shall not resort to cycle averaging or postulates that total work done by pulsation should equal kinetic energy change, as these are applicable in a limited scope only to nearly-adiabatic pulsation. Instead we use \mEq{WSQ} and rearrange it to obtain the balance relation, including the norm, coupling and boundary terms. This gives an exact identity within the adopted linear pulsation system, rather than an approximate cycle averaged energy argument.

We start by expressing volume perturbation ($q$) by using continuity equation derived in Appendix. From \mEq{ContQ} volume perturbation is given by
\[
q=d'+3d-\Lambda h.
\]

Substituting this equation into \mEq{WSQ} it is obtained that
\begin{equation}
\begin{aligned}
W_s&=-\int_{x_b}^{x_s} C\Re(p\,\overline{\sigma d'}) dx -3\int_{x_b}^{x_s} C\Re(p\,\overline{\sigma d}) dx \\
   &+\Lambda\int_{x_b}^{x_s} C\Re(p\,\overline{\sigma h}) dx.
\end{aligned}
\label{eq:WSDP}
\end{equation}
Integrating the first term in \mEq{WSDP} by parts leads to
\[
W_s=-\Re\left[Cp\,\overline{\sigma d}\right]_{x_b}^{x_s}+\Re\int_{x_b}^{x_s}\left[(Cp)'-3Cp\right]\overline{\sigma d}\,dx+\Lambda\int_{x_b}^{x_s}C\Re(p\,\overline{\sigma h})dx.
\]
In Zalewski (2026c) it was shown that the term $(Cp)'-3Cp=C(p'-A_3p)$, thus the above formula may be written as
\[
W_s=-\Re\left[Cp\,\overline{\sigma d}\right]_{x_b}^{x_s}+\Re\int_{x_b}^{x_s}\left[C(p'-A_3p)\right]\overline{\sigma d}\,dx+\Lambda\int_{x_b}^{x_s}C\Re(p\,\overline{\sigma h})dx.
\]
Substituting \mEq{RM}, the derivative of pressure is eliminated leading to 
\begin{equation}
	\begin{aligned}
W_s&=-\Re\left[Cp\,\overline{\sigma d}\right]_{x_b}^{x_s} \\
&+\Re\int_{x_b}^{x_s}\left[CA_3\left((4-A_2\sigma^2-A_5)d-w_1-\Lambda  h\right)\right]\overline{\sigma d}\,dx \\
&+\Lambda\int_{x_b}^{x_s}C\Re(p\,\overline{\sigma h})dx.
    \end{aligned}
\label{eq:BalanceWS}
\end{equation}

\section{The generalized balance relation}

\subsection{Coupling terms}

The \mEq{BalanceWS} can be simplified by introducing surface terms, norm integral $\NS$ and coupling terms as
\begin{equation}
\WS+B_{\rm p, surf}-B_{\rm p,bot}=\RES\NS+C_{\rm ph}+C_{\rm hd}+C_{\rm g},
\label{eq:BalanceEQ}
\end{equation}
where the surface terms are given by
\[
B_{\rm p} = C\Re(p\,\overline{\sigma d}),
\]
and the radial norm integral is
\[
\NS=\NC+\NT,
\]
where
\[
\begin{aligned}
		\NC &= -\int_{x_b}^{x_s}C A_3A_2|\sigma|^2|d|^2\,dx \\
        \NT &= \int_{x_b}^{x_s}C A_3(4-A_5)|d|^2\,dx
\end{aligned}
\]
and the coupling terms are given by
\[
\begin{aligned}
	C_{\rm hd}&=-\Lambda\int_{x_b}^{x_s}C A_3\Re(h\,\overline{\sigma d})\,dx, \\
    C_{\rm ph}&=\Lambda\int_{x_b}^{x_s}C\Re(p\,\overline{\sigma h})\,dx \\
    C_{\rm g}&=-\int_{x_b}^{x_s}C A_3\Re(w_1\,\overline{\sigma d})\,dx
\end{aligned}
\]

\subsection{Reduction of coupling terms}
The coupling terms in \mEq{BalanceEQ} may be simplified with the use of horizontal momentum \mEq{HM} to write
\[
C_{\rm ph}+C_{\rm hd}=\RES\left(N_{\rm in,h}+N_{\rm dh}\right)+C_{\rm gh},
\]
where
\[
\begin{aligned}
	N_{\rm in,h}&=-\Lambda\int C A_3 A_2 |\sigma|^2 |h|^2\, dx,\\
	N_{\rm dh}&=-2\Lambda\int C A_3 \Re(d\,\overline{h})\,dx,
\end{aligned}
\]
and
\[
C_{\rm gh}=-\Lambda\int C A_3 \Re(w\,\overline{\sigma h}) dx.
\]

The term $C_{\rm gh}$ may be combined with $C_{\rm g}$ to obtain
\[
C_{\rm g}+C_{\rm gh}=-\int C A_3 \Re\left[ w_1\,\overline{\sigma d}+\Lambda w\,\overline{\sigma h}\right]\,dx.
\]

Introducing complex gravitational functional
\[
{\cal J}_{\rm g}=\int C A_3 (w_1\,\overline{d}+\Lambda w\,\overline{h})\, dx,
\]
the above may be rewritten introducing
\[
N_{\rm g}=-\Re({\cal J}_{\rm g}),
\]
as
\[
C_{\rm g}+C_{\rm gh}=-\Re(\overline{\sigma}{\cal J}_{\rm g})=\RES N_{\rm g}-\Im(\sigma)\Im({\cal J}_{\rm g}).
\]

By adopting the derivation presented in Reese et al. (2021) the last term may be transformed using Green's identity together with Poisson equation to obtain, \mEq{JgBoundary},
\[
\Im({\cal J}_{\rm g})=\Im[{\cal G}(x_{\rm s})-{\cal G}(x_{\rm b})],
\]
where ${\cal G}$ is given by
\[
{\cal G}=C A_3\left(w\,\overline{d}+\frac{w\,\overline{w_1}}{A_5}\right),
\]
and it is composed of two terms - viz. matter - displacement, and gravitational field terms.

Hence the last term in equation for $C_{\rm g}+C_{\rm gh}$ is a difference of boundary terms and should be included in $B$ as
\[
B=B_{\rm p}+B_{\rm g},
\]
with
\[
B_{\rm g}=\Im(\sigma)\Im({\cal G}).
\]                   

The terms can be further simplified by noticing that $N_{\rm mech}$ can be combined with $N_{\rm dh}$ by introducing the radial amplitude $\epsilon_{\rm A}$ of the tangential-area perturbation $\epsilon_{\rm A}(r)Y^m_\ell$ 
\[
\epsilon_{\rm A}=2\,d-\Lambda\,h.
\]
The sum of $N_{\rm mech}+N_{\rm dh}$ can be written using $\epsilon_{\rm A}$ as
\[
\begin{aligned}
N_{\rm mech}+N_{\rm dh}&=\int C A_3 \Re{\left[(4-A_5)|d|^2-2\Lambda d\,\overline{h}\right]}\, dx \\
&=2\int C A_3 \Re(d\,\overline{\epsilon_{\rm A}})\,dx-\int C A_3 A_5 |d|^2\,dx.		
\end{aligned}
\]
Introducing
\[
\begin{aligned}
	N_{\rm ha}&=2\int C A_3 \Re(d\,\overline{\epsilon_{\rm A}})\, dx,\\
		N_5&=-\int C A_3 A_5 |d|^2\,dx,
\end{aligned}
\]
it is obtained that
\[
N_{\rm mech}+N_{\rm dh}=N_{\rm ha}+N_5.
\]
The last term can be combined with $N_{\rm g}$ to form the gravitational and stratification norm term
\[
N_{\rm sg}=N_{\rm g}+N_{\rm 5}.
\]

\subsection{The balance equation}
Using the above formulae the balance equation may be written as
\begin{equation}
W_{\rm s}+\PHS-\PHB=\RES(N_{\rm in}+N_{\rm ha}+N_{\rm sg}),
\label{eq:BalanceTot}
\end{equation}
with the combined, radial and horizontal, inertia term
\[
N_{\rm in}=\NC+N_{\rm in,h}=-\int_{x_{\rm b}}^{x_{\rm s}} C A_3 A_2 |\sigma|^2\left(|d|^2+\Lambda |h|^2\right)\, dx.
\]

\subsection{Kinetic Energy - Power relation}
\label{sec:KEPR}

The term $N_{\rm in}$ may be written in terms of luminosity scaled mode inertia $\mathcal{I}$ and $|\sigma|^2$ (Christensen-Dalsgaard 2014, Unno et al. 1989) as
\[
N_{\rm in}=-|\sigma|^2\mathcal{I}=-2\EKIN,
\]
where 
\[
\mathcal{I}=\int_{x_{\rm b}}^{x_{\rm s}} C A_3 A_2 \left(|d|^2+\Lambda|h|^2\right)\,dx,
\]
and mode kinetic energy scaled by luminosity is given by
\begin{equation}
\EKIN=\frac{1}{2}|\sigma|^2\mathcal{I}>0.
\label{eq:EKIN}
\end{equation}

It is thus possible to rewrite the balance equation for linear nonadiabatic pulsation, both radial and nonradial, as
\begin{equation}
	\PTOT+\PDB=2\RES\EKIN,
\label{eq:EKED}	
\end{equation}	
with
\[
\PTOT=\PTOT_{\rm t}+\PTOT_{\rm c}+\PTOT_{\rm ha}+\PTOT_{\rm sg},
\]
and
\[
\PDB=B_{\rm bot}-B_{\rm surf}
\]
where
\[
\begin{aligned}
	\PTOT_{\rm t}    &= -\WT, \\
	\PTOT_{\rm c}    &= \RES N_{\rm c}, \\
	\PTOT_{\rm ha} &= \RES N_{\rm ha}, \\
	\PTOT_{\rm sg}   &= \RES N_{\rm sg}.
\end{aligned}
\]
and where $\PTOT_{\rm c}$ was obtained from $\WC$, since the $\WC$ term in \mEq{WS} may be written as 
\[
\WC=-\RES N_{\rm c},
\]
with
\[
N_{\rm c}=-\int_{x_{\rm b}}^{x_{\rm s}} C A_4 |p|^2\,dx.
\]

The above \mEq{EKED} represents kinetic energy - power balance for a mode for linear nonadiabatic pulsation, both radial or nonradial. It retains the positive-definite kinetic energy term, and separates volume-power terms from boundary terms. The latter contain both pressure-work and gravitational terms. 

For radial pulsation $h=0$, hence $N_{\rm ha}=4\int C A_3|d|^2dx$. In this limit $N_{\rm g}=-N_5$, and therefore $N_{\rm sg}=0$. Thus, for radial pulsation, $N_{\rm ha}=N_{\rm str}$ is the structural norm introduced for radial pulsations in Zalewski (2026c) and the terms in \mEq{EKED} reduce to
\[
\begin{aligned}
	\PEX&=\PTOT_{\rm t}=-W_{\rm t},\\
	\PRSP&=\RES\left(N_{\rm c}+N_{\rm str}\right),\\
	&=-W_{\rm c}+\RES N_{\rm str}
\end{aligned}
\]
and $2\EKIN=-N_{\rm in}$.
The surface terms reduce to $B=B_{\rm p}$ and the balance relation may be written as
\[
W_{\rm s}-\PDB_{\rm p}=\RES(N_{\rm in,r}+N_{\rm str}).
\]

For nonradial pulsation in the Cowling approximation $N_{\rm g}=0$, so that $N_{\rm sg}=N_5$; this correction is small in the low-density envelope layers where $A_5\ll 1$.

\subsection{Exchange and response power terms}
\label{sec:ERPT}

The Kinetic energy - Power relation \mEq{EKED} may be written as
\begin{equation}
\PEX+\PRSP+\PDB=2\RES\EKIN,
\label{eq:PowerEk}
\end{equation}
where 
\[
\begin{aligned}
\PEX&=\PTOT_{\rm t}, \\
\PRSP&=\PTOT_{\rm c}+\PTOT_{\rm ha}+\PTOT_{\rm sg}.
\end{aligned}
\]
The symbol $\PEX$ denotes the energy exchange terms, while $\PRSP$ collects the terms that provide response contribution. The \mEq{PowerEk} may be viewed as
\[
\PEX+\PDB=2\RES\EKIN-\PRSP,
\]
and since the response terms are $\sim\RES$ it may be written as
\[
\PEX+\PDB=-\RES N_{\rm eff},
\]
where
\[
N_{\rm eff}=N_{\rm in}+N_{\rm c}+N_{\rm ha}+N_{\rm sg}.
\]

This form of the balance equation shows that the power is generated by the exchange and boundary terms. The terms in $N_{\rm eff}$ do not constitute independent sources of driving or damping. Instead, they modify the proportionality between the exchange power and the amplitude growth rate. The effective norm, $N_{\rm eff}$, is not sign definite. Consequently the sign of $\PEX+\PDB$ alone does not, in general, determine the sign of $\RES$. 

The successive forms of the balance relation are retained because they expose different aspects of the same identity: the coupling structure, the effective norm, and the relation between total power and positive-definite kinetic energy.

The role of the $\PTOT$-terms in \mEq{EKED} for various types of envelope modes will be examined in the subsequent section. 

\subsection{Growth-rate diagnostics}
\label{sec:GRDG}

Since for a given mode the quantities entering the balance relation may be computed, including the surface terms, it is possible to use any of the forms of the balance equation - viz. \mEq{BalanceEQ}, \mEq{BalanceTot} or \mEq{EKED} to compute a check, $\RES_{\rm check}$, of the real part of eigenfrequency to verify the accuracy of the BVP solution
\begin{equation}
\RES_{\rm check}=\frac{\PTOT+\PDB}{2\EKIN}.
\label{eq:ReCheck}
\end{equation}
This check is applicable to near adiabatic as well as strongly nonadiabatic radial and nonradial modes.

For nearly adiabatic pulsation an equation of the form of \mEq{ReCheck} is sometimes used to obtain an estimate of mode excitation. However since $\PTOT$ contains the response terms which explicitly depend on $\RES$ it is better to estimate $\RES_{\rm est}$ using \mEq{BalanceTot} to obtain
\begin{equation}
	\RES_{\rm est}=\frac{\PEX+\PDB}{2\EKIN-N_{\rm rsp}},
\label{eq:ReEst}	
\end{equation}
with $N_{\rm rsp}=N_{\rm c}+N_{\rm ha}+N_{\rm sg}$.
This form of the balance equation, while more suitable for the estimation of the excitation rate requires the evaluation of the response terms in addition to inertia and power terms. Since even for radial mode the response terms $N_{\rm c}$ and $N_{\rm ha}$ do not vanish the denominator may not be positive-definite.

In order to use \mEq{ReEst} for radial modes in addition to the usual eigenfunctions an adiabatic pressure perturbation would be required to compute $N_{\rm c}$, while for nonradial modes knowledge of several other perturbations - viz. horizontal displacement, as well as gravitational potential perturbations would be required.

\section{Application to envelope modes}
In what follows a  model of the AGB envelope for $M=0.69M_{\odot}$,$L=10^4 L_{\odot}$, $\log(T_{\rm eff})=3.8$ is used in the calculations. For the different types of envelope modes examined the boundary-condition selectors are chosen to minimize the surface terms $B$ and result in well behaved modes (see Zalewski 2026a). Typically the selectors $(3,4)\text{--}(1,3)$ are used.

The use of Cowling approximation is adequate for the nonradial p-mode pulsations in 
extended AGB envelopes considered here, as the envelope mass is small
compared with the stellar mass, and for which $A_5\ll 1$ in the
outer layers. 

Since only envelope models are used, the discussion of gravity
modes is necessarily restricted to modes whose amplitudes are
concentrated in the outer envelope. In particular, we consider
low-frequency g-like modes which carry most of their energy near
the H/He~I ionization zones, although their propagation
region may extend deeper into the envelope.

For nonradial modes the inner boundary is not placed at a region where $\tau_{th}/\tau_{ac}\sim 1$ as is done for radial modes. Instead we truncate
the mode according to the Lamb-frequency criterion. The boundary is
placed at the first layer satisfying
\begin{equation}
	\frac{\ell(\ell+1)}{A_2A_3A_4|\sigma|^2}\ge RC2^2 ,
	\label{eq:LambRC}
\end{equation}
which corresponds to $S_{\ell}/|\omega|\ge RC2$ in the adopted
normalization. The parameter $RC2$ is chosen empirically; in the
calculations discussed below we use $RC2$ in the range $1$--$100$.

The choice of the type of inner boundary selector is particularly important
for nonradial modes. This differs from the radial case, where the inner boundary can
usually be placed deep enough that the pulsation variables have
already decayed. For nonradial envelope modes the criterion
\mEq{LambRC} often places the boundary much higher in the
envelope, where the solution may still be oscillatory. For nonradial modes we use at the inner boundary, either the selector $(1,3)$,
which admits a fast branch of the local dispersion relation and a
slow branch, or a selector $(3,4)$ which admits two slow branches at the inner boundary, or $(1,2)$ which admits two fast branches, depending on the mode type. The notation and branch selection were described in Zalewski (2026a).

In the computations presented here the selector
$(1,3)$ gives regular amplitudes near the inner boundary for nonradial strange modes and low-frequency g-modes, while the selector $(3,4)$ is better suited to ordinary nonradial p-modes for which the inner boundary is placed much deeper. The inner boundary selector of $(1,2)$ is used for radial modes and for low-frequency nonradial thermal modes (see Zalewski (2026a) for discussion of inner boundary selectors for radial pulsation).

A mode is
accepted only if the balance integrals stabilize before the inner
boundary is reached, the amplitude remains regular, and the
spillover coefficients at the boundary (Zalewski 2026b) are small. We also use the boundary terms ($B$) to assess the choice of boundary conditions and select the boundary-condition selectors such that these terms do not make a significant contribution to the balance relation. While the $\PDB$ terms are much smaller than the power terms and can be neglected in the analysis of the role of power term components, the boundary terms are needed to obtain proper balance.

For nonradial p-modes with frequencies corresponding to $\sim 3$-rd overtone or above and for the envelope model considered here the singular-value maps (see Zalewski 2026b) are smooth. At lower frequencies, and especially
for the g-like modes in the envelope, the maps cease to be smooth and show additional structure. This behavior appears to be associated with changes in the character of the local dispersion branches at the frequency-dependent inner boundary location.

Pulsation equations were integrated using the continuous renormalization method augmented by the tracking transformation described in Zalewski (2026b). For each mode reported below, we have verified that its eigenfrequency and eigenfunction morphology were locally robust with respect to variations of RC2 about the adopted value.

In what follows we will omit from the discussion the boundary terms ($\PDB$) as, due to the choice of boundary conditions, their values are negligible compared to the power terms in \mEq{PowerEk} for the analyzed modes.

\subsection{Ordinary p-modes}
From the analysis performed for radial modes in AGB envelopes, it was found  (Zalewski 2026c) that for ordinary p-modes
\[
\begin{aligned}
|N_{\rm str}|&\gg |N_{\rm in,r}|, \\
|W_{\rm t}|&\ll |W_{\rm c}|,
\end{aligned}
\]
and thus
\[
\begin{aligned}
W_{\rm s}&\sim W_{\rm c}, \\
N_{\rm s}&\sim N_{\rm str}.
\end{aligned}
\]

Hence for ordinary p-modes the balance is determined not by the thermodynamic driving term $W_{\rm t}$ and mode inertia ($N_{\rm in,r}$) but by the terms $W_{\rm c}$ and $N_{\rm str}$ not present in the classical derivation based on dissipation and kinetic energy integrals.

\begin{figure}[htb]
	\includegraphics{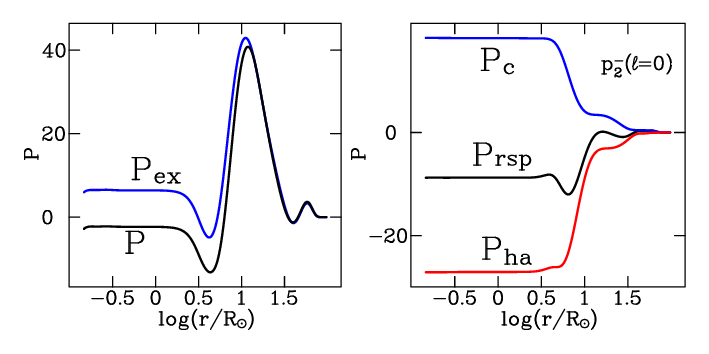}
	\FigCap{2-ov radial, damped mode ($p^-_2$). The total, luminosity normalized, power $\PTOT$ compared to the driving $\PEX$ is shown in the left panel. Luminosity normalized response power $\PRSP$ and its components $\PTOT_{\rm c}$ and $\PTOT_{\rm ha}$ are shown in the right panel. For radial pulsation the term $\PTOT_{\rm sg}=0$. The exchange power term $\PEX>0$ but total power, $\PTOT<0$.}
	\label{fig:Fig1}
\end{figure} 

The role of the response terms, \mEq{PowerEk}, in the power balance may be seen by plotting the integral of the luminosity normalized mode exchange power $\PEX=-W_{\rm t}$ and comparing it with the luminosity normalized net power $\PTOT=\PEX+\PRSP$. The results for a 2-ov are shown in \mFig{Fig1}. 

From the right panel of \mFig{Fig1} it may be seen that the area deformation power $\PTOT_{\rm ha}$ makes substantial contribution to $\PRSP$, opposing the compression power term $\PTOT_{\rm c}$, and it is the power loss on the area deformation that makes the $\PRSP<0$.

It follows from \mFig{Fig1} that the driving $\PEX>0$ for this mode, whereas the response contribution has larger and opposite value. Consequently the total power is negative and the computed mode is stable. This shows that the response power terms are not insignificant and can affect mode excitation rate, and it is not always sufficient for the driving term to be positive for the mode to be unstable.

\begin{figure}[htb]
	\includegraphics{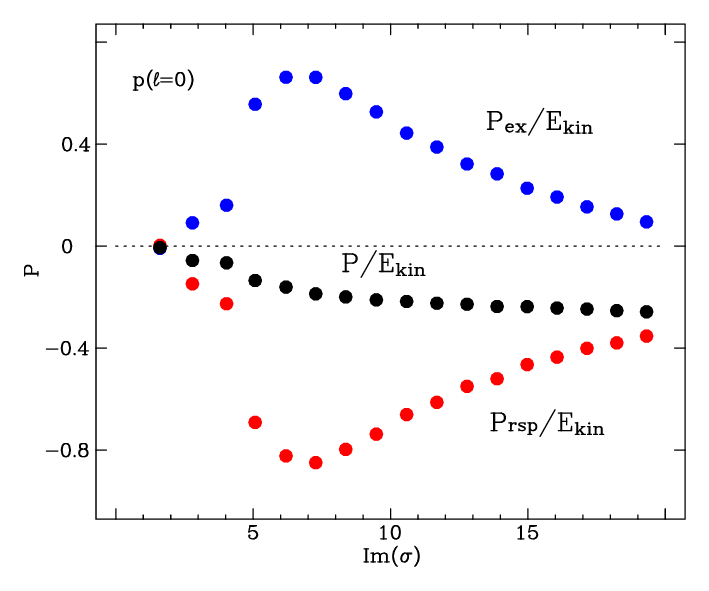}
	\FigCap{The power $\PTOT$ and its components $\PEX$ and $\PRSP$ normalized by mode kinetic energy for a sequence of ordinary radial ($\ell=0$) p-modes, from F to 16-ov. The response term outweighs the driving term and the total power is negative.}
	\label{fig:Fig2}
\end{figure}

In \mFig{Fig2} the total power $\PTOT$ and the driving and response components are plotted for a sequence of radial ordinary p-modes ranging from fundamental to 16-ov. It may be seen that the response term opposes the mode driving supplied by the exchange term, $\PTOT<0$, even though $\PEX>0$. 

This means that the radial p-modes for this model are not stable because of the lack of driving, but because the driving is insufficient to overcome the negative response term.

\begin{figure}[htb]
	\includegraphics{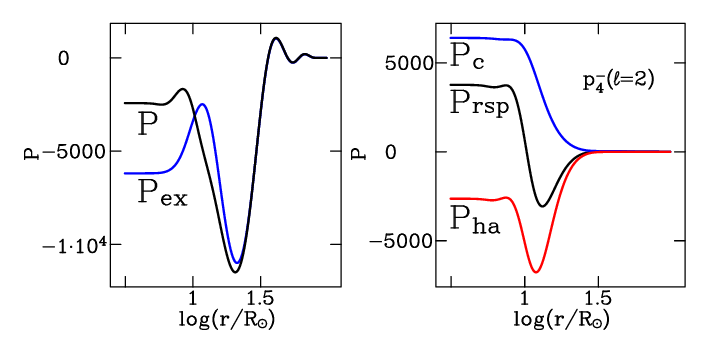}
	\FigCap{The plot of total power $\PTOT$ and the exchange power $\PEX$ for a damped, nonradial ($\ell=2$), 4-ov p-mode is shown in the left panel. The response power together with its components is shown in the right panel. Since $|\PTOT_{\rm ha}|<|\PTOT_{\rm c}|$ the $\PRSP>0$ and thus $\PTOT>\PEX$.}
	\label{fig:Fig4}
\end{figure} 

The behavior of the components of $\PTOT$ for a nonradial p-mode is presented in \mFig{Fig4} for a 4-ov damped mode for $\ell=2$ ($p_4^-$). For the $p_4^-(\ell=2)$ mode the term $\PEX<0$ thus it contributes to mode damping, but the response term $\PRSP$ is of opposite sign to $\PEX$, and is partially offsetting the exchange power, yet the total power is negative and the mode is stable. 

\begin{figure}[htb]
	\includegraphics{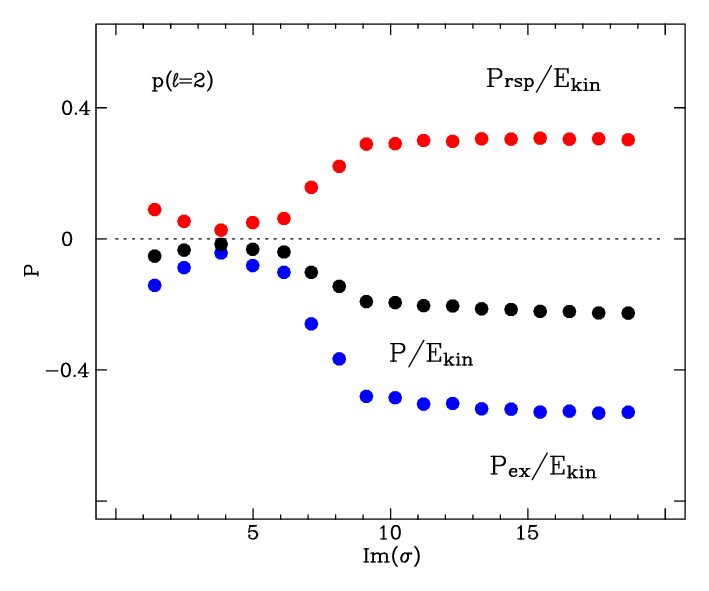}
	\FigCap{The $\PTOT$ and its components for a sequence of $\ell=2$ p-modes. The exchange power $\PEX<0$ and $|\PEX|>|\PRSP|$ thus the $\ell=2$ p-modes for this model are damped.}
	\label{fig:Fig5}
\end{figure}

The same occurs for other ($\ell=2$) p-modes for this model as may be seen from \mFig{Fig5} where the power and its components are shown for a sequence of modes.

While for nonradial $\ell=2$ p-modes the signs of the exchange and response terms are reversed compared to radial p-modes of the same model, the contribution of the response term to stability is substantial in both cases.

Glasner and Buchler (1993) analyzed the dependence of excitation rate on frequency for radial modes in RR Lyr and Cepheid models. Our results for $\PEX$ for radial modes in AGB stars (\mFig{Fig2}) also exhibit maximum of driving (around 4-5 overtone modes), but the negative response has larger magnitude which makes the total power $\PTOT<0$ and the total power decreases with the increase of frequency and does not show maxima. This indicates that for both radial and nonradial modes such analyses should include not only the thermodynamic work integral but also the coupling terms to obtain excitation rates.

\begin{figure}[htb]
	\includegraphics{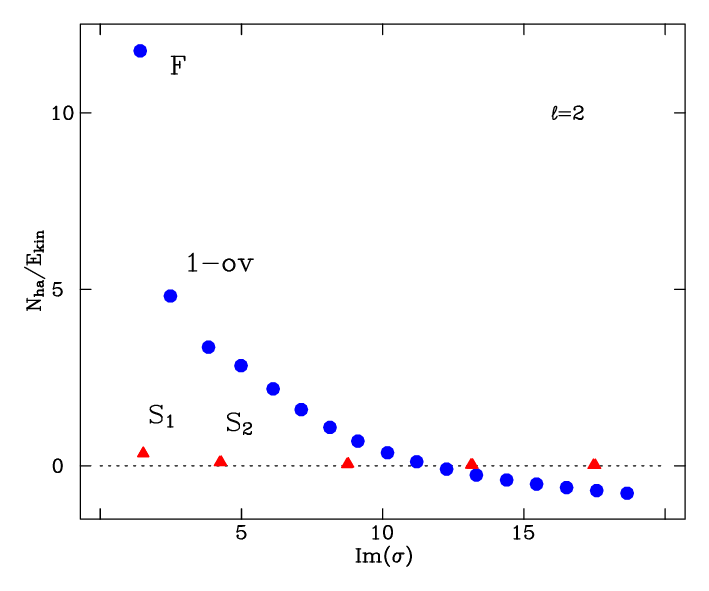}
	\FigCap{Normalized area-deformation norm $N_{\rm ha}/\EKIN$ for a sequence of nonradial ($\ell=2$) p-modes (blue dots) and strange modes (red triangles) is shown for a range of frequencies. For p-modes the $N_{\rm ha}>0$ below 10-ov for this model. For the nonradial strange modes the $N_{\rm ha}$ is small and not significant compared to the compression term. }
	\label{fig:Fig6}
\end{figure}

The compression power term $\PTOT_{\rm c}$ entering the response power $\PRSP$ is $\PTOT_{\rm c}=\RES N_{\rm c}$, and the integrand of $N_{\rm c}$ is negative definite. Thus the sign of $\PTOT_{\rm c}$ is always $-\sgn(\RES)$. The area-deformation power $\PTOT_{\rm ha}$ integrand depends on the real part of the product of radial displacement ($d$) and amplitude of tangential area perturbation ($\epsilon_{\rm A}$) and thus may make the $\PTOT_{\rm ha}$ depend not only on the sign of the real part of pulsation frequency but also on the relative amplitudes of radial and horizontal displacement entering $\epsilon_{\rm A}$. In \mFig{Fig6} a plot of the area-deformation norm $N_{\rm ha}$ normalized by mode kinetic energy ($\EKIN$) is presented for a sequence of both nonradial ($\ell=2$) p-modes and strange modes for the same model as in \mFig{Fig5}.

From \mFig{Fig6} it may be seen that the area-deformation power $\PTOT_{\rm ha}$ term changes sign for higher overtone nonradial p-modes. Thus for p-modes with the same sign of $\RES$ but differing in frequency it may have an opposite effect on the total power. For this sequence of damped nonradial p-modes, the compression term opposes the term $\PEX$. 

\subsection{Strange p-modes}
In Zalewski (2026c) it was found that the work and norm integrals for radial strange modes differ substantially from those for ordinary modes. This is reexamined here using present formulation based on power and kinetic energy rate for radial and for nonradial strange modes. 

\begin{figure}[htb]
	\includegraphics{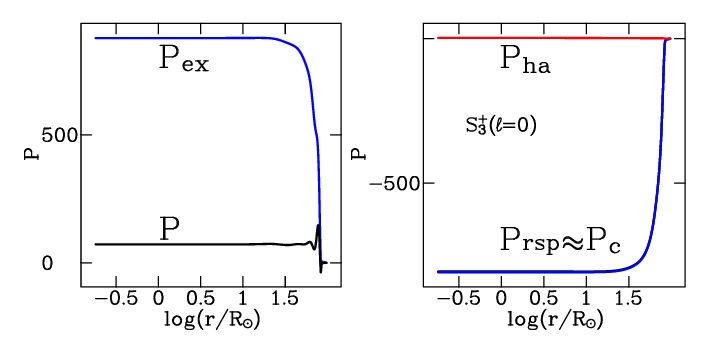}
	\FigCap{Strange $S_3^+$ radial mode. The total power $\PTOT$ compared to the driving $\PEX$ is shown in the left panel. Response power $\PRSP$ and its components $\PTOT_{\rm c}$ and $\PTOT_{\rm ha}$ are shown in the right panel. For this mode $\PTOT>0$ but it is substantially reduced by $\PRSP$.}
	\label{fig:Fig3}
\end{figure}

\begin{figure}[htb]
	\includegraphics{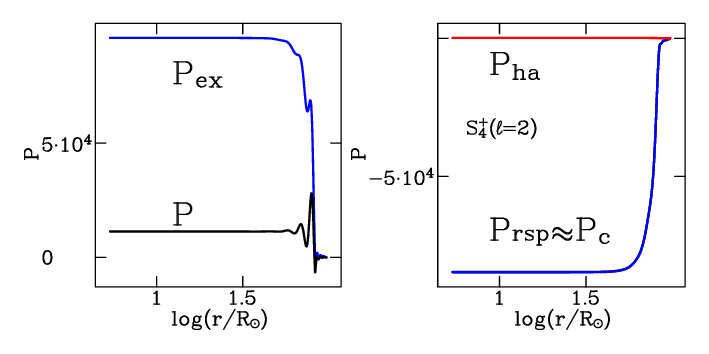}
	\FigCap{The power $\PTOT$ and exchange power $\PEX$ for nonradial ($\ell=2$) strange mode $S^+_4$ are shown in the left panel. The components of the response power $\PRSP$ terms are shown in the right panel. The response term decreases substantially the power that can be used to change kinetic energy of the mode.}
	\label{fig:Fig7}
\end{figure}

In \mFig{Fig3} the power $\PTOT$ and its components are displayed for a $S^+_3$ radial strange mode and in \mFig{Fig7} for a $S^+_4$ nonradial ($\ell=2$) strange mode.

The properties of $\PEX$ and $\PRSP$ for the radial and nonradial strange modes are essentially the same for both considered excited strange modes. In both cases the area deformation power $\PTOT_{\rm ha}$ is small compared to the compression power term $\PTOT_{\rm c}$, thus the response power is nearly equal to the compression power. 

In Zalewski (2026c) it was shown that $W_{\rm c}\sim-W_{\rm t}$ for strange radial modes. In the present notation this can be stated as $\PRSP\sim-\PEX$ because of small $\PTOT_{\rm ha}$. Thus the total power for a strange mode is much smaller than any of the excitation or compression terms.

\begin{figure}[htb]
	\includegraphics{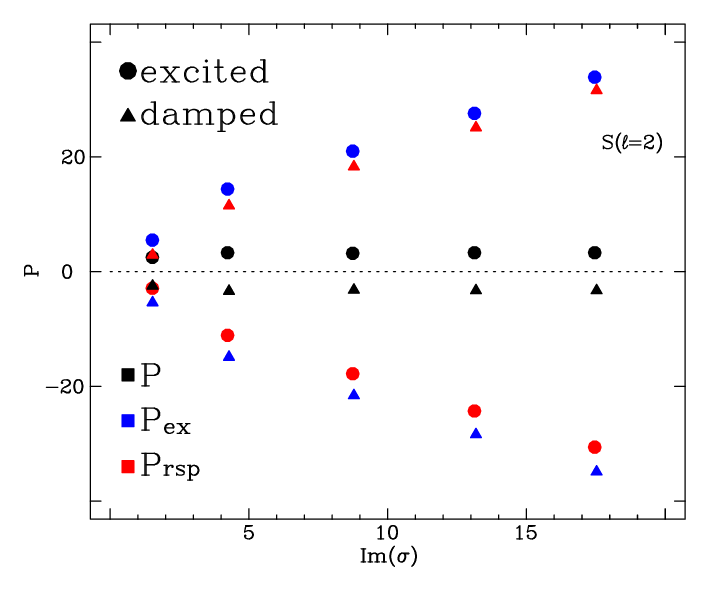}
	\FigCap{A composite diagram showing the excitation and response power terms and the resulting total power for a sequence of five pairs of strange, nonradial ($\ell=2$) modes. The exchange power term of a strange mode is nearly compensated by the compression term.}
	\label{fig:Fig8}
\end{figure}

The total power and the $\PEX$ and $\PRSP$ for a sequence of nonradial excited and damped strange modes for $\ell=2$ are shown in \mFig{Fig8}. A similar diagram, but using work terms $W$ and for radial strange modes was presented in Zalewski (2026c, Fig.~5). In both cases the structure of the diagram is the same - viz. a particular excited strange mode has small but positive value of $\PTOT$ while the $\PEX>0$ is large and $\PRSP\sim-\PEX$. The two large terms nearly cancel out. For the damped strange mode in the pair with $\sigma_\pm=(\pm \gamma,\nu)$ the signs of the $\PEX$ and $\PRSP$ terms, and thus of the total power are reversed.

\subsection{Low frequency outer envelope modes}
\begin{figure}[htb]
	\includegraphics{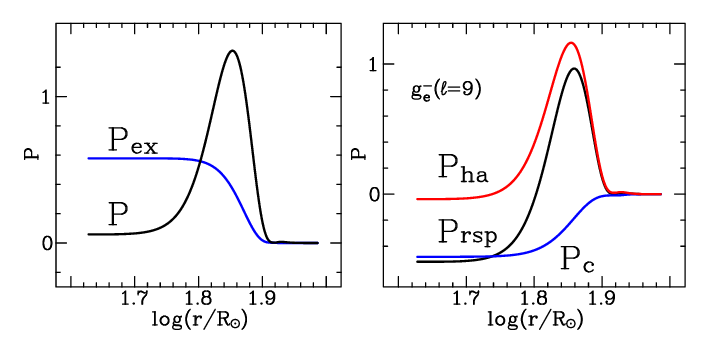}
	\FigCap{The total power and the exchange power are shown in the left panel for an excited $g^-_e$ mode. The components of the response power are shown in the right panel. The area-deformation power $\PTOT_{\rm ha}$ peaks in the H/He~I ionization region, but overall effect of this term is small compared to the compression term. The total power of the mode is substantially reduced by the response term.}
	\label{fig:Fig9}
\end{figure}

Since we are using unfitted envelopes we are capable of finding only outer envelope g-modes. In \mFig{Fig9} we present the power terms for a nonradial ($\ell=9$), low frequency $\sigma=(0.15, 0.31)$ excited g-mode trapped near the H/He~I ionization zone. The mode extends from surface to a region above the He~II ionization. Its main amplitude and kinetic energy are located in the convection region just below the H/He~I ionization zone. This mode is accompanied by a damped mode with $\sigma=(-0.15, 0.31)$. The two seem to form a pair of g-modes, possibly of the $g^{-}$ type described by Saio (2011) and Saio et al. (2015). The excited mode of the pair is shown in \mFig{Fig9} and is provisionally denoted as $g^-_e(\ell=9)$.

It is seen from this figure that for the $g^-$-like mode both the $\PTOT_{\rm ha}$ and $\PTOT_{\rm c}$ are negative, contrary to ordinary p-modes. The area-deformation power term has a small magnitude, thus $\PRSP\sim\PTOT_{\rm c}$. The response term reduces substantially the effect of the excitation term $\PEX$, however the mode remains unstable ($\PTOT>0$).

It is interesting that for both of these $g^-$-like modes the $\arg(p\,\overline{s})\approx 0$ in the region $1.75\le \log(r/R_{\odot})\le 1.91$. Hence the phase relation between pressure and entropy perturbation is for these modes the same as for strange modes discussed in Zalewski (2026c). It thus seems that this particular type of low frequency surface g-modes exhibits properties reminiscent of strange modes, at least by the phase $p$-$s$ relation and the opposite-sign terms $\PTOT_{\rm c}$ and $\PEX$.

\subsection{Thermal modes}
Thermal modes associated with the structure of the linearized thermal-diffusion operator and their influence on radial stellar pulsations were studied by Pesnell and Buchler (1986). Here we consider a thermal eigenmode of the full nonradial boundary-value problem primarily to examine how its terms enter the amplitude-level power balance.

\begin{figure}[htb]
	\includegraphics{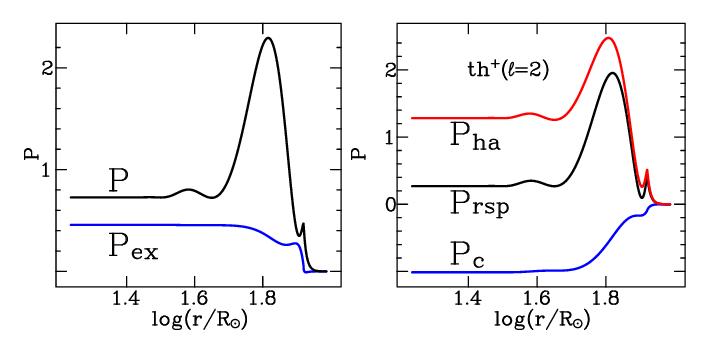}
	\FigCap{The power and the excitation term are shown in the left panel for a thermal ($th^+(\ell=2)$) excited nonradial mode. The area-deformation and the compression terms oppose each other leading to small, however positive, value for the response term.}
	\label{fig:Fig10}
\end{figure}

The behavior of the power terms for a thermal nonradial mode ($th^+(\ell=2)$) is shown in \mFig{Fig10}. The frequency of this mode is $\sigma=(0.28, 0)$. There also exists a damped mode with $\sigma=(-0.28,0)$. 

For the nonradial excited thermal mode the total power is larger than the exchange power $\PTOT>\PEX$ because the response term $\PRSP$ is positive. The area-deformation power term $\PTOT_{\rm ha}$ outweighs the compression term $\PTOT_{\rm c}$ making the response power positive.

Although the thermal mode is non-oscillatory, $\Im(\sigma)=0$, the formulation presented in Section~\ref{sec:KEPR} yields a finite kinetic-energy amplitude because the displacement still changes exponentially when $\RES\ne 0$. Specifically,
\[
\EKIN=\frac{1}{2}|\sigma|^2\mathcal{I}
\]
depends on the modulus of the full complex eigenfrequency rather than only on its oscillatory part, $\Im(\sigma)$.

\section{Conclusions}
Using the postulated sesquilinear form representing an analogue to $-PdV$ pressure work (\mEq{KS}), and adopting the approach outlined by Glatzel (1994) to express the work  in terms of pulsation variables using momentum and continuity equations, we have derived an equation analogous to Glatzel Eq.~(6.5) but applicable to nonadiabatic, nonradial pulsation using our variables and exposing the constituent terms (Zalewski 2026c). These terms are the work integrals $W=\WC+\WT$ and the norm integrals, as shown in \mEq{BalanceTot}.

The balance equation \mEq{BalanceTot} is applicable to nonadiabatic, nonradial pulsation and relates work done and surface boundary terms to the rate of change of the norm integral. The norm integral in case of nonradial, nonadiabatic pulsation is composed of several terms. Due to this the norm integral is not sign definite, contrary to the classical formulation which relies on positive-definite kinetic energy $\EKIN=-N_{\rm in}/2$. 

The \mEq{BalanceTot} may be used as a balance relation, for example to check the computed eigenfrequency or to obtain an estimate of excitation rate provided the norm terms and work done are computed based on a suitable approximation of the eigenfunctions, as discussed in Section~\ref{sec:GRDG}

Because the norm $N_{\rm eff}$ is not sign definite it is preferable to recast the balance relation into a form reminiscent of classical work done-energy rate form as derived in \mEq{EKED} in Section~\ref{sec:KEPR} We show that it is possible to write the balance relation in a form relating the rate of change of kinetic energy ($\EKIN$) and total power ($\PTOT$) of the mode including the surface terms ($\PDB$)
\[
\PTOT+\PDB=2\RES\EKIN.
\]

The power $\PTOT$ is composed, for nonadiabatic, nonradial pulsations, of two principal terms - viz. 
the exchange power ($\PEX$) obtained from the usual thermodynamic term 
$\PTOT_{\rm t}$, which depends on pressure ($p$) and entropy ($s$) perturbations via 
$p\,\overline{\sigma s}$, and the response power term $\PRSP$ which combines terms that represent 
the compression, the horizontal area distortion and gravitational field distortion power terms, 
as discussed in Section~\ref{sec:ERPT}

These response power terms are all proportional to $\RES$ and would vanish for a neutral mode ($\RES=0$). 
Ignoring the boundary terms, the thermodynamic term is the only volume term representing energy exchange. The response terms are not independent sources or sinks, but they affect the relation between the exchange power and the amplitude growth rate through the effective norm.

The terms entering the response represent the reaction to the driving. However, not all of them are sign definite when real part of the corresponding sesquilinear form is taken. The only explicitly sign-definite norm terms are 
$N_{\rm in}$ - which is used to define kinetic energy of pulsation, and $N_{\rm c}$ - the compressional norm which is 
used to define $\PTOT_{\rm c}$. The remaining - area-deformation and gravitational-stratification terms 
are not manifestly sign-definite, as has been shown for the $N_{\rm ha}$ term for a 
sequence of nonradial ($\ell=2$) p-modes in \mFig{Fig6}, and thus their effect on mode excitation rate may vary.

It was found for the radial modes that the combined effect of the response power terms may result 
in the radial modes being stable ($\PTOT<0$) even though the driving power terms for them ($\PEX$) are positive. This is seen in  \mFig{Fig2}, particularly for higher overtone modes. 
This shows that the power available for a mode may be distributed not entirely to the change of mode kinetic energy 
but also into the other response terms. A mode becomes unstable only if $\mathcal{P}+\Delta B>0$. 
For nonradial ($\ell=2$) p-modes the exchange term 
$\PEX<0$, but response $\PRSP>0$. This shows that if the nonradial p-modes are damped the response terms counterbalance the damping, but not sufficiently to make the modes unstable.

A useful conclusion follows from \mEq{PowerEk} - viz. at the edge of instability strip for a particular mode, where $\RES=0$ the response power term $\PRSP=\RES N_{\rm rsp}=0$. The right hand side of the equation becomes zero. This leads to a relation between the thermodynamic power $\PEX$ and the boundary terms difference $\PDB$
\[
(\PEX+\PDB)|_{\RES=0}=0,
\]
where we apply this relation at the blue edge of the mode's instability strip as the effects of convection on the balance of power, important near the red edge (see Houdek and Dupret 2015), are not included in our derivation. 

This relation implies that at the blue edge the thermodynamic power term may be non-zero only due to the surface terms. Thus it is in principle possible to shift the edge by assuming forms of boundary conditions that would lead to substantial role of surface terms. The solution of the boundary value problem would then adapt to imposed boundary conditions, altering the eigenfunctions and thus affecting the $\PEX$ power term. In the present article boundary conditions were chosen so as to minimize the contribution from $\PDB$.

Thus the response terms do not enter the neutral-stability condition explicitly, although their norm contributions may affect the behavior of the growth rate on either side of the boundary.

The relation between power terms for the case of strange modes differs from ordinary p-modes in that the horizontal area-distortion power term is small, much smaller than the compression term (see eg. \mFig{Fig3}), so that the $\PRSP\approx \PTOT_{\rm c}$. The response power nearly balances the exchange power term making the total power $\PTOT$ small. This effect occurs both for the examined radial (Zalewski 2026c) and nonradial ($\ell=2$) strange modes and may be seen for a sequence of nonradial strange modes in \mFig{Fig8}. For strange modes thus the balance is obtained by the interplay of the exchange and compression power terms.

The analyzed low-frequency, outer envelope pair of gravity modes with frequencies 
$\sigma_{\pm} \approx (\pm 0.15, 0.31)$ have much in common with the oscillatory convection modes $g^-$ described by Saio (2011). In both cases the kinetic energy is localized in the convection zone, in the present case, just below the H/He~I ionization region. Additionally we have found that the power terms $\PEX$ and $\PTOT_{\rm c}$ are of opposite signs, while the $\PTOT_{\rm ha}$ term is small. Also for these two modes the pressure and entropy perturbations in the region of maximum energy exchange are in phase such that $\arg(p\,\overline{s})\approx 0$. In this respect they exhibit properties similar to strange modes (Zalewski 2026c).
 
We have also found that the balance relation between the power of the mode and its rate of change of kinetic energy holds for thermal modes (for which $\Im(\sigma)=0$) provided kinetic energy is defined as in \mEq{EKIN}.

The numerical applications presented here were restricted to envelope models and evaluated in the Cowling approximation for nonradial modes. Consequently, the gravitational-stratification power contribution $\PTOT_{\rm sg}$ could not be examined in its general non-Cowling form.

\appendix
\renewcommand{\theequation}{A\arabic{equation}}
\setcounter{equation}{0}

\section{Linearized equations for mechanical and gravitational-potential perturbations}
To derive a balance equation applicable to both radial and nonradial pulsations we will need linearized continuity, momentum and Poisson's equations. We will write those equations in a form suitable for present purpose by following Dziembowski (1971, 1977) derivation and using the same notation.

\subsection{Continuity equation}
The linearized continuity equation may be written as
\[
\frac{\partial\delta\rho}{\partial t}+\boldsymbol{\nabla}\cdot\left(\rho\frac{\partial \boldsymbol{\xi}}{\partial t}\right)=0,
\]
where $\delta$ is Eulerian perturbation and $\boldsymbol{\xi}$ is the displacement perturbation
\[
\boldsymbol{\xi}={\xi}_r(r)Y^m_\ell\boldsymbol{e}_r+\xi_h(r)\boldsymbol{\nabla}_hY^m_\ell,
\]
where $\boldsymbol{e}_r$ is the radial unit vector, and $Y^m_\ell$ is the spherical harmonic of degree $\ell$ and azimuthal order $m$. It is convenient to introduce nondimensional radial and horizontal displacement perturbations as
\[
d=\frac{\xi_r}{r},\quad h=\frac{\xi_h}{r}.
\]
By introducing $\exp(\omega t)$ time dependence of perturbations and Lagrangian perturbation of density $\Delta \rho$ the linearized continuity equation may be rewritten as
\[
\frac{\Delta\rho}{\rho}+\boldsymbol{\nabla}\cdot\boldsymbol{\xi}=0,
\]
where the divergence term written as
\[
\boldsymbol{\nabla}\cdot\boldsymbol{\xi}=\left(d'+3\,d-\ell(\ell+1)h\right)Y^m_\ell,
\]
where $'$ denotes differentiation with respect to the independent variable $x=\ln(r/R_{\odot})$.
In what follows we will need the relative volume perturbation, hence we introduce
\[
q=\frac{\Delta V}{V}=-\frac{\Delta\rho}{\rho},
\]
and write the continuity equation as
\begin{equation}
	q=d'+3d-\Lambda h, \quad \Lambda=\ell(\ell+1).
	\label{eq:ContQ}
\end{equation}

\subsection{Gravitational-potential perturbations}
The Eulerian perturbation of the Newtonian gravitational potential satisfies Poisson's equation
\[
\nabla^2\delta\Phi=4\pi G\delta\rho.
\]
where the angular dependence of perturbation variables is
\[
\delta\Phi(r,\theta,\phi)=\delta\Phi(r)Y^m_\ell(\theta,\phi).
\]
By introducing
\[
w=\frac{r}{GM_r}\delta\Phi
\]
and
\[
w_1=\frac{r^2}{GM_r}\frac{d\delta\Phi}{dr}
\]
the quantity $w$ may be related to $w_1$ as
\[
w_1=\frac{r^2}{GM_r}\frac{d}{dr}\left(\frac{GM_r}{r}w\right).
\]
By introducing 
\[
A_5=\frac{d\ln M_r}{d\ln r}=\frac{4\pi r^3\rho}{M_r}
\]
this equation may be written as
\begin{equation}
	w'=w(1-A_5)+w_1
\label{eq:EW}
\end{equation}

The Poisson equation may be written as
\[
\frac{1}{r^2}\frac{d}{dr}\left(r^2\frac{d\delta\Phi}{dr}\right)-\frac{\ell(\ell+1)}{r^2}\delta\Phi=4\pi G\delta\rho.
\]
Using $w$ and $w_1$ and multiplying by $r^3/(GM_r)$ this equation may be rewritten as
\[
\frac{r}{M_r}\frac{d}{dr}(M_rw_1)-\Lambda w=\frac{4\pi r^3\rho}{M_r}\frac{\delta\rho}{\rho}.
\]
Using $A_5$, the derivative of $M_rw_1$ may be expressed as
\[
\frac{r}{M_r}\frac{d}{dr}(M_rw_1)=w'_1+A_5w_1,
\]
so finally
\[
w'_1=A_5\left(\frac{\delta\rho}{\rho}-w_1\right)+\Lambda w.
\]
By using
\[
A_3=-\frac{d\ln P}{d\ln r},
\]
and
\[
A_4=\frac{1}{\Gamma_1}
\]
with
\[
A_6 = \frac{1}{\Gamma_1}\frac{d\ln P}{d\ln r}-\frac{d\ln\rho}{d\ln r} \]
the density derivative is expressed as
\[
\frac{d\ln\rho}{d\ln r}=-A_6-A_3A_4,
\]
hence it is obtained that
\begin{equation}
	w'_1=A_5\left[d\left(A_6+A_3A_4\right)+\frac{\Delta\rho}{\rho}-w_1\right]+\Lambda w.
\label{eq:EW1}
\end{equation}

\subsection{Horizontal component of momentum equation}
The linearized momentum equation may be written as
\[
\frac{\partial^2\boldsymbol{\xi}}{\partial t^2}=-\frac{1}{\rho}\boldsymbol{\nabla}\delta P-\frac{\delta\rho}{\rho}\boldsymbol{\nabla}\Phi-\boldsymbol{\nabla}\delta\Phi.
\]
By introducing
\[
p_{\rm e}=\frac{\delta P}{P},\quad p=\frac{\Delta P}{P}
\]
and using $A_3$ one may write
\[
p_{\rm e}=p+A_3d.
\]
For a scalar perturbation $F(r)Y^m_\ell$,
\[
\boldsymbol{\nabla}\left(FY^m_\ell\right)=\frac{dF}{dr}Y^m_\ell\boldsymbol{e}_r+\frac{F}{r}\boldsymbol{\nabla}_hY^m_\ell.
\]
Therefore the linearized momentum equation may be separated into a radial component, proportional to $Y^m_\ell\boldsymbol{e}_r$, and a horizontal component, proportional to $\boldsymbol{\nabla}_hY^m_\ell$. 

Using this the horizontal component of momentum equation may be written as
\[
\omega^2r^2h=-\frac{P}{\rho}p_{\rm e}-\delta\Phi.
\]
Since
\[ \frac{P}{\rho}=\frac{gr}{A_3}=\frac{GM_r}{rA_3}
\]
and using $w$ the horizontal part of linearized momentum equation may be written as
\[
\frac{\omega^2r^3}{GM_r}h=-\left(\frac{p_{\rm e}}{A_3}+w\right).
\]
By introducing nondimensional frequency $\sigma$ given by
\[
\omega=\sqrt{4\pi G\langle\rho\rangle}\sigma
\]
and a coefficient
\[
 A_2=\frac{4\pi r^3\langle\rho\rangle}{M_r}
\]
 the equation may be rewritten as
\[
h=-\frac{1}{A_2\sigma^2}\left(d+\frac{p}{A_3}+w\right).
\]
and introducing 
\[
\eta=\frac{\Lambda}{A_2\sigma^2}
\] 
it may be rewritten as
\begin{equation}
	\Lambda h=-\eta\left(d+\frac{p}{A_3}+w\right).
	\label{eq:HM}	
\end{equation}

\subsection{Radial component of momentum equation}
The radial part of the momentum equation may be written as
\[
\omega^2rd=-\frac{1}{\rho}\frac{d\delta P}{dr}-\frac{\delta\rho}{\rho}g-\frac{d\delta\Phi}{dr}.
\]
The last term may be written as $gw_1$. Rearranging the radial momentum equation it is obtained that
\[
p'_{\rm e}=A_3\left[p_{\rm e}-\frac{\delta\rho}{\rho}-w_1-A_2\sigma^2d\right].
\]
The derivative of $p_{\rm e}$ may be written as
\[
p'_{\rm e}=p'+A'_3d+A_3d'
\]
and the Eulerian density perturbation may be written as
\[
\frac{\delta\rho}{\rho}=\frac{\Delta\rho}{\rho}-d\frac{d\ln\rho}{d\ln r}.
\]
Substituting these quantities into the radial momentum equation and using the linearized continuity equation \mEq{ContQ} to eliminate $d'$ it is obtained
\[
p'=A_3\left(p-w_1-A_2\sigma^2d-\Lambda h+d\left[A_3+\frac{d\ln\rho}{d\ln r}+3-\frac{A'_3}{A_3}\right]\right).
\]
Since
\[
g=\frac{GM_r}{r^2},
\]
we have
\[
\frac{d\ln g}{d\ln r}=A_5-2.
\]
Since 
\[
A_3=\frac{\rho gr}{P}=-\frac{d\ln P}{d\ln r},
\]
then
\[
\frac{A'_3}{A_3}=\frac{d\ln\rho}{d\ln r}+\frac{d\ln g}{d\ln r}+1-\frac{d\ln P}{d\ln r}.
\]
Hence
\[
A_3+\frac{d\ln\rho}{d\ln r}+3-\frac{A'_3}{A_3}=4-A_5.
\]
Substituting the radial part of linearized momentum equation becomes
\begin{equation}
	p'=A_3\left[\left(4-A_2\sigma^2-A_5\right)d+p-w_1-\Lambda h\right].
\label{eq:RM}
\end{equation}
By using \mEq{HM} this equation may be further brought to the form of Eq.~(2) in Dziembowski (1977) by introducing
\[
\eta_1=\frac{\eta}{A_3}.
\]

The Lagrangian density perturbation may be expressed in terms of pressure ($p$) and entropy ($s$) perturbations using coefficients $A_4$ and 
\[
A_7=\left(\frac{\partial\ln \rho}{\partial\ln T}\right)_{\!P}
\]
as
\[
\frac{\Delta\rho}{\rho}=A_4p+A_7s
\]
with
\[
s=\frac{\Delta S}{c_P}=\frac{\Delta T}{T}-\nabla_{ad}\frac{\Delta P}{P}=t-A_8A_{10}p.
\]

\renewcommand{\theequation}{B\arabic{equation}}
\setcounter{equation}{0}

\section{Reduction of the gravitational functional to boundary terms}
\label{app:gravboundary}

Following Reese et al. (2021), we translate the Green-identity reduction of the gravitational coupling functional into the variables and normalization used here. We show that the imaginary part of the gravitational functional
introduced in Section~5 is determined entirely by its values at the
boundaries. Let
\[
\phi(r)\equiv\delta\Phi(r)
\]
denote the radial amplitude of the gravitational-potential
perturbation, so that the complete perturbation is
$\phi(r)Y_\ell^m(\theta,\varphi)$. We adopt the angular normalization
implicit in the factor $4\pi$ entering $C(x)$,
\[
\int |Y_\ell^m|^2\,d\Omega=4\pi,
\qquad
\int \boldsymbol{\nabla}_{h}Y_\ell^m\cdot
\boldsymbol{\nabla}_{h}\overline{Y_\ell^m}\,d\Omega
=4\pi\Lambda ,
\]
where $\Lambda=\ell(\ell+1)$.

Consider the complex gravitational coupling functional
\begin{equation}
	{\cal I}_{\rm g}
	=
	\int_V \rho\,\overline{\boldsymbol{\xi}}\cdot
	\boldsymbol{\nabla}\delta\Phi\,dV .
	\label{eq:Ig3D}
\end{equation}
Using
\[
\boldsymbol{\xi}
=r dY_\ell^m\boldsymbol{e}_r
+r h\boldsymbol{\nabla}_{h}Y_\ell^m,
\]
the angular integration gives
\begin{equation}
	{\cal I}_{\rm g}
	=
	4\pi\int_{r_b}^{r_s}\rho
	\left(
	r^3\frac{d\phi}{dr}\,\overline{d}
	+\Lambda r^2\phi\,\overline{h}
	\right)dr.
	\label{eq:IgRadial}
\end{equation}

The linearized continuity equation for the Eulerian density
perturbation may be written as
\[
\delta\rho=-\boldsymbol{\nabla}\cdot(\rho\boldsymbol{\xi}).
\]
Using
\[
\boldsymbol{\nabla}\cdot
\left(\rho\delta\Phi\,\overline{\boldsymbol{\xi}}\right)
=
\rho\,\overline{\boldsymbol{\xi}}\cdot
\boldsymbol{\nabla}\delta\Phi
+\delta\Phi\,
\boldsymbol{\nabla}\cdot
\left(\rho\overline{\boldsymbol{\xi}}\right),
\]
Eq.~(\ref{eq:Ig3D}) becomes
\[
{\cal I}_{\rm g}
=
\oint_{\partial V}
\rho\delta\Phi\,
\overline{\boldsymbol{\xi}}\cdot\boldsymbol{n}\,dS
+\int_V\delta\Phi\,\overline{\delta\rho}\,dV.
\]
The complex conjugate of Poisson's equation gives
\[
\nabla^2\overline{\delta\Phi}
=4\pi G\overline{\delta\rho}.
\]
Consequently,
\[
{\cal I}_{\rm g}
=
\oint_{\partial V}
\rho\delta\Phi\,
\overline{\boldsymbol{\xi}}\cdot\boldsymbol{n}\,dS
+\frac{1}{4\pi G}
\int_V\delta\Phi\,
\nabla^2\overline{\delta\Phi}\,dV.
\]
Applying Green's first identity to the last integral yields
\begin{equation}
	\begin{aligned}
		{\cal I}_{\rm g}
		={}&
		\oint_{\partial V}
		\left[
		\rho\delta\Phi\,
		\overline{\boldsymbol{\xi}}\cdot\boldsymbol{n}
		+\frac{1}{4\pi G}\delta\Phi\,
		\frac{\partial\overline{\delta\Phi}}{\partial n}
		\right]dS
		\\
		&-\frac{1}{4\pi G}
		\int_V
		\boldsymbol{\nabla}\delta\Phi\cdot
		\boldsymbol{\nabla}\overline{\delta\Phi}\,dV .
	\end{aligned}
	\label{eq:IgGreen}
\end{equation}
The volume integral in Eq.~(\ref{eq:IgGreen}) is real because
\[
\boldsymbol{\nabla}\delta\Phi\cdot
\boldsymbol{\nabla}\overline{\delta\Phi}
=
|\boldsymbol{\nabla}\delta\Phi|^2.
\]
It follows that the imaginary part of ${\cal I}_{\rm g}$ is supplied
only by the boundary terms:
\begin{equation}
	\Im({\cal I}_{\rm g})
	=
	\Im\oint_{\partial V}
	\left[
	\rho\delta\Phi\,
	\overline{\boldsymbol{\xi}}\cdot\boldsymbol{n}
	+\frac{1}{4\pi G}\delta\Phi\,
	\frac{\partial\overline{\delta\Phi}}{\partial n}
	\right]dS .
	\label{eq:IgImaginary}
\end{equation}

For a spherical shell bounded by $r_b$ and $r_s$, the outward normal
is $\boldsymbol{e}_r$ at $r_s$ and $-\boldsymbol{e}_r$ at $r_b$.
After angular integration the local boundary expression is
\begin{equation}
	{\cal I}_{\partial}(r)
	=
	4\pi\rho r^3\phi\,\overline{d}
	+\frac{r^2}{G}\phi\,
	\frac{d\overline{\phi}}{dr}.
	\label{eq:IgSurfacePhysical}
\end{equation}
Hence
\[
\Im({\cal I}_{\rm g})
=
\Im\left[
{\cal I}_{\partial}(r_s)
-{\cal I}_{\partial}(r_b)
\right].
\]

The relation between ${\cal I}_{\rm g}$ and the dimensionless
functional used in Section~5 follows from
\[
\Omega_0=\sqrt{4\pi G\langle\rho\rangle},
\qquad
dx=\frac{dr}{r},
\]
and the definitions of $C$, $A_3$, $A_5$, $w$, and $w_1$. Direct
substitution into complex gravitational functional $\mathcal{J}_{g}$ and using \mEq{IgRadial} leads to
\begin{equation}
	{\cal J}_{\rm g}
	=
	\int_{x_b}^{x_s}CA_3
	\left(
	w_1\overline{d}+\Lambda w\overline{h}
	\right)dx
	=
	\frac{\Omega_0}{2L}{\cal I}_{\rm g}.
	\label{eq:JgIg}
\end{equation}

Using the same definitions of $C$, $A_3$, $A_5$, $w$, and $w_1$, the corresponding dimensionless local boundary expression is
\begin{equation}
	{\cal G}(x)
	=
	CA_3
	\left(
	w\overline{d}
	+\frac{w\overline{w_1}}{A_5}
	\right)
	=
	\frac{\Omega_0}{2L}{\cal I}_{\partial}(r).
	\label{eq:GgPhysical}
\end{equation}
Substitution of \mEq{GgPhysical} into the preceding relation finally gives
\begin{equation}
	\Im({\cal J}_{\rm g})
	=
	\Im\left[
	{\cal G}(x_s)-{\cal G}(x_b)
	\right].
	\label{eq:JgBoundary}
\end{equation}

\end{document}